\documentclass{aa}  

\usepackage{graphicx}
\usepackage{txfonts}
\begin{document}

   \title{An investigation of the low-$\Delta$V near-Earth asteroids (341843)~2008~EV5 and (52381)~1993~HA
   \thanks{Based on observations collected at the
          European Organisation for Astronomical Research in the Southern Hemisphere
          under ESO programmes 290.C-5013 and 095.C-0087,
          and at the Observat\'orio Astron\^omico do Sert\~ao de Itaparica
          of the Observat\'orio Nacional, Brazil.}}

   \subtitle{Two suitable targets for the ARM and MarcoPolo-M5 space missions}

\titlerunning{An investigation of the low-$\Delta$V NEAs 2008~EV5 and 1993~HA}

   \author{D. Perna
          \inst{1}
          \and
          M. Popescu\inst{1,2}
          \and
          F. Monteiro\inst{3}
          \and
          C. Lantz\inst{1}
          \and
          D. Lazzaro\inst{3}          
          \and
          F. Merlin\inst{1}
          }

   \institute{LESIA -- Observatoire de Paris, PSL Research University,
CNRS, Sorbonne Universit\'es, UPMC Univ. Paris 06, Univ. Paris Diderot,
Sorbonne Paris Cit\'e, 5 place Jules Janssen, 92195 Meudon, France\\
              \email{davide.perna@obspm.fr}
         \and
             Astronomical Institute of the Romanian Academy,
             5 Cuţitul de Argint, 040557 Bucharest, Romania
         \and
         Observatório Nacional, R. Gal. José Cristino 77, 20921-400, Rio de Janeiro, Brazil
             }

   \date{Received 10 June 2016; accepted 14 September 2016}

 
  \abstract
   {The Asteroid Redirect Mission (ARM) under development by NASA
   is being planned to collect a multi-meter boulder from a near-Earth asteroid (NEA),
   and to bring it to the cis-lunar space in the mid-2020's for
   future study and exploitation by a crewed mission.
   The MarcoPolo-M5 project is being proposed in 2016 for the M5 mission opportunity by ESA,
   to bring back to Earth a sample from a very primitive D-type NEA.
   As D-types are very rare within the NEA population, considerable effort is still in progress
   to characterize easily accessible targets with unknown surface composition, in order to
   discover further asteroids that belong to this taxonomic group.
   }
   {We aim to further characterize the physical properties of two
   optimal targets for sample return space missions, the low-$\Delta$V NEAs (341843) 2008 EV5
   and (52381) 1993 HA. The asteroid 2008 EV5 is
   the baseline target of ARM, but only one spectrum of this object exists in the literature.
   The asteroid 1993 HA is a very favourable target for a space mission based on
   its dynamical properties: here we intend to assess if it is a suitable target for MarcoPolo-M5.
   }
   {We obtained visible spectroscopy of 2008 EV5 with the FORS2 instrument at ESO-VLT (Paranal, Chile),
   at different rotational phases.
   We also obtained visible and near-infrared spectroscopy of 1993 HA, using the
   EFOSC2 and SOfI instruments at ESO-NTT (La Silla, Chile). Visible photometry of 1993 HA was
   carried out within the IMPACTON project at the
   Observat\'orio Astron\^omico do Sert\~ao de Itaparica (Itacuruba, Brazil). 
   }
   {
   Our new observations are in agreement with the C-type classification of 2008 EV5,
   which is a requirement for the ARM mission.
   We obtained five visible spectra which
   do not show any variability within the limits of noise, suggesting a homogeneous surface.
   We obtained the first ever spectroscopic dataset ($\sim$0.4-1.6 $\mu$m) for 1993 HA, finding
   a featureless, red-sloped behaviour typical of D-types
   (a T or X classification is also possible, with decreasing confidence).
   We also found that the synodic rotation period of 1993 HA is 4.107$\pm$0.002 h,
   a value that is optimal for the execution of a sample return mission.
   The derived lightcurve also suggests an elongated shape (axis ratio a/b$\ge$1.71).   
   At this stage 1993 HA does indeed seem to be the most favourable target for MarcoPolo-M5,
   though future observations are necessary to study it further.
   }
   {}


   \maketitle
%

\section{Introduction}
Over recent years, many scientific and technological goals have pushed space
agencies to launch -- and plan for the near future --
space missions to near-Earth asteroids (NEAs; see, e.g. Barucci et al. 2011).
However, due to the wide variety of orbital and physical characteristics of NEAs,
target selection must be able to guarantee both technical feasibility and high scientific return
(e.g., Perozzi et al. 2001; Binzel et al. 2004a; Ieva et al. 2014).
The accessibility of potential targets of space
missions from Earth is studied by classical orbital transfer algorithms:
for example, the $\Delta$V parameter (i.e. the necessary velocity change  applied
to a spacecraft to realise a rendez-vous mission, following the approach
by Shoemaker \& Helin 1978) can be used to identify the ``easiest'' targets to reach.
From a scientific point of view, the primitive, carbonaceous asteroids
appear to be the most appealing targets: they contain very pristine material formed in the outer solar system
which can provide fundamental information on the origin and early evolution of the solar system itself
(e.g. Barucci et al. 2012),
and they are likely to represent a resource of water and rare minerals
to be exploited by humanity in the near future (e.g. Sanchez \& McInnes 2013).
Given the low number of known objects with such characteristics,
the search for further low-$\Delta$V NEAs of primitive nature has become
a major topic in planetary science.

The Asteroid Redirect Mission (ARM; Abell et al. 2016) is being developed by NASA
to visit an NEA with a robotic mission
(to be launched in December 2021 on the current schedule),
collect a multi-ton boulder from its surface,
and redirect it into a stable orbit around the Moon.
A second crewed mission (to be launched in December 2026)
will explore the boulder and return to Earth with samples.
The ARM mission intends to test a number of capabilities needed for future human missions,
as well as the gravity tractor technique (Lu \& Love 2005) to deflect the target asteroid orbit.
The baseline target of ARM is NEA (341843) 2008 EV5,
chosen because of its carbonaceous (C-type) nature and
short ($\sim$4.5 years) round-trip mission scenario.
Indeed, the only available spectrum of 2008 EV5 is typical
of primitive C-type asteroids and shows a weak ($\sim$3\%) spectral absorption at 0.48 $\mu$m,
suggesting the presence of alteration minerals
with similarities to the CI primitive meteorite Orgueil (Reddy et al. 2012).
The albedo of 2008 EV5 (0.12$\pm$0.04, Busch et al. 2011)
is also consistent with a C-type taxonomic classification.

A primitive asteroid sample return mission has been long studied at ESA in the framework of the 
Cosmic Vision Programme.
Hayabusa 2 (launched in December 2014 by JAXA) 
and OSIRIS-REx (launched in September 2016 by NASA) will return samples of
a C- and a B-type asteroid, respectively, to Earth. 
Following these missions,
the new proposal
MarcoPolo-M5 (proposed in 2016 for the M5 mission opportunity of
ESA's Cosmic Vision Programme) will study
a very primitive, organic-rich D-type NEA (supposed to have originated in the outer solar system),
and return bulk samples to Earth for laboratory analyses.
This investigation will allow us to better understand the origin of planetary materials and
to identify and characterize the organics and volatiles in a primitive asteroid.
This will enable us to check
the exobiological scenario for the
origin of life which invokes an exogeneous delivery to the early Earth of complex organic molecules
by primitive bodies, capable of triggering the pre-biotic synthesis of biochemical compounds on our planet.
As with ARM, the target selection procedure for MarcoPolo-M5 is fundamental in the mission planning,
and involves both the dynamical and physical study of all the 
favourable targets
(we note that D-type objects are very rare in the NEA population;
see, e.g. Binzel et al. 2004b and Perna et al. 2016).
The baseline target for the MarcoPolo-2D proposal
(not selected in 2015 for the Cosmic Vision M4 opportunity
because of the M4 budgetary restrictions)
was identified in the D-type 2001 SG286 (e.g., Popescu et al. 2011).

In this work we present new spectral observations of 2008 EV5,
to confirm that it is indeed a good candidate target for the NASA ARM mission,
as well as the first ever spectral data of (52381) 1993 HA:
this asteroid offers a very efficient operational and technical mission profile,
with a complete mission scenario of only 3.6 years
(instead of the five or six years required for 2001 SG286)
and could be
an optimal target for MarcoPolo-M5.
Further photometric observations of 1993 HA
allowed us to retrieve its rotation period and constraints on its shape.
It is of utmost importance that the rotation period
of the target chosen by a sample return mission
should not be shorter than $\sim$2~h, so as to be confident that
a regolith is present on the surface and to have realistic expectations
concerning the outcome of sampling operations.
Furthermore, the rotation period needs to be fast enough
to facilitate global mapping (e.g., Barucci et al. 2009).

\begin{table*}[t]
\caption{Observational circumstances for spectroscopic data.}
\label{obs}      
\centering
\small{
\begin{tabular}{cccccccc}        
\hline\hline                 
Object   & Telescope-Instrument & Date     & UT$_{start}$ & t$_{exp}$ (s) & Airmass   & Solar analog (airmass) & Rotational phase\\
\hline                        

2008 EV5 & VLT-FORS2     & 13 Mar 2013 & 9:29         & 1440          & 2.03-1.74 & SA 102-1081 (2.44) & 0\\
         &               & 15 Mar 2013 & 9:25         & 1440          & 2.12-1.81 & SA 112-1333 (2.13) & 0.87\\
         &               & 16 Mar 2013 & 9:18         & 1600          & 2.25-1.87 & SA 112-1333 (2.10) & 0.28\\
         &               & 17 Mar 2013 & 9:15         & 1800          & 2.33-1.88 & SA 102-1081 (2.18) & 0.72\\
         &               & 22 Mar 2013 & 9:43         & 1000          & 1.96-1.77 & SA 112-1333 (1.68) & 0.03\\
\hline
1993 HA  & NTT-EFOSC2    & 5 Nov 2015  & 6:31         & 2$\times$450  & 1.06      & HD 11123 (1.02)    & 0\\
         &               & 15 Dec 2015 & 6:01         & 600           & 1.03      & SA 98-978 (1.14)   & 0.63$\pm$0.11\\
         & NTT-SOfI      & 15 Dec 2015 & 7:00         & 14$\times$120 & 1.10-1.16 & HIP 27185 (1.12)   & 0.87$\pm$0.11\\          
\hline
\end{tabular}}
\end{table*}
%

\section{Observations and data reduction}
Visible spectroscopy of 2008 EV5 was obtained at the ESO Very Large Telescope (Paranal, Chile)
using the FORS2 (Appenzeller et al. 1998) instrument and the 150I grism.
Spectroscopic observations of 1993 HA were carried out at the ESO New Technology Telescope (La Silla, Chile)
using the EFOSC2 instrument (Buzzoni et al. 1984) for the visible range (Grism 1)
and the SOfI instrument (Moorwood et al. 1998) for the near-infrared (NIR)
range (blue grism).
The observational circumstances are given in Table~1. 
All the observations have been performed orienting the slit (2\arcsec) along the parallactic angle, to minimize the effects of atmospheric differential refraction.
The nodding technique of moving the object along the slit between two different positions
was used for NIR observations,
as is necessary at these wavelengths for a proper background subtraction.

Data were reduced using standard procedures
(bias and background subtraction, flat field correction, one-dimensional spectra extraction,
atmospheric extinction correction -- see Perna et al. 2015 for more details)
with MIDAS and IRAF packages.
Wavelength calibration was obtained using emission lines
from the lamps available at each instrument and which cover the spectral intervals considered.
The reflectivity of our targets was then obtained by dividing their visible and NIR spectra by those of
solar analogs observed close in time and in airmass to the scientific frames (cf. Table 1).

Photometric data of asteroid 1993 HA were acquired within the
IMPACTON project (Lazzaro 2010) at the Observat\'orio Astron\^omico do Sert\~ao de
Itaparica (OASI,
Itacuruba, Brazil) on two nights, the 7 and 9 December 2015.
The observations, in
the R band, were
made using a $2048 \times 2048$ Apogee Alta U42 CCD camera, giving a $11.8
\times 11.8$~arcmin
field. 
The asteroid was observed for nearly 7.5 h using an exposure
time
of 120 s and starting at about 23:20 UT in each one of the two observing
nights. During this time the airmass of the target decreased from $\sim$2 to
$\sim$1.07,
then increased again up to $\sim$1.7 at the beginning of the morning
twilight,
when we stopped our observations. At the time of the observations the
asteroid
distance to Sun and Earth was 1.095 AU and 0.176 AU, respectively,
and
the solar phase was 47.4 degrees.

Data reduction was performed using the MaxIm DL package, following the
standard procedures
of flat-field correction and sky subtraction, and the observation time was
corrected for light-travel time.


\section{Data analysis and discussion}

\subsection{2008 EV5}
We obtained five different spectra of 2008 EV5.
The observing conditions were far from ideal, as the asteroid was observable only during
the astronomical twilight, at high airmass (cf. Table~1),
and at a relatively faint magnitude (V=22.2).
Thus the resulting signal-to-noise ratio (S/N) of the reduced spectra was very low,
ranging between $\la$1.5 (for the spectra acquired on 15 and 22 March 2013)
and approximately three (spectrum acquired on 17 March 2013).
Hence we rebinned the spectra by replacing 10-pixel intervals by their median value,
in order to increase such low S/N.
Figure~1
shows the final spectra, where the S/N was improved by a factor of approximately three.
The rotational phases for each spectrum are also reported,
based on the rotational period of 3.725$\pm$0.001~h measured by Galad et al. (2009).

Despite the overall low quality of our data, they are important
as
the next
observability window for 2008 EV5 is not until December 2020,
too late for proper planning of the ARM mission.
The spectra are too noisy to confirm or deny the presence of the weak 
0.48 $\mu$m feature observed by Reddy et al. (2012),
as well as to perform a proper taxonomic classification. However
these new observations of the 400-m large (Mainzer et al. 2011) 2008 EV5
suggest a flat and featureless spectrum (within the limits of noise)
in agreement with the C-type classification indicated
by the literature (a visible and near-infrared spectrum by Reddy et al. 2012,
visible photometry by Somers et al. 2010 and Ye 2011).
No evident variability appears with the rotational phase,
providing the first evidence
for 2008 EV5's homogeneous surface.

\subsection{1993 HA}
We obtained good photometric and spectroscopic data for 1993 HA, thanks
to its relatively bright magnitude at the time of our observations (V$\sim$18.2)
and favourable sky conditions.

Two individual lightcurves
were obtained over the two nights of observation
via the calculation of relative magnitudes, that is, the difference between the
instrumental
magnitude of the asteroid and that of one field comparison star with similar
magnitude.
This was done in the interests of minimizing the effects of the atmospheric
extinction and
weather changes, resulting in a mean error of about 0.015 mag. 
These time-series were
modelled with a 4th order
Fourier polynomial (Harris et al. 1989), whose best fit relation
corresponded to a synodic
rotation period of 4.107$\pm$0.002~h. The data folded with the best
fit are given in Fig.~2. 
The double-peaked composite lightcurve is well covered with data from the two nights,
and has an amplitude of 0.58 mag.
From the relation $\Delta$m=2.5log(a/b),
where $\Delta$m is the maximum lightcurve amplitude reached in equatorial view,
we estimated a lower limit to the axis ratio
of the triaxial ellipsoid shape of a/b$\ge$1.71, implying a quite elongated body.
We note that such elongation does not set any particular constraint
on the execution of a sample return mission. It is worth noting that the NEA (25143) Itokawa,
target of the Hayabusa sample return mission, is even more elongated
(e.g., Kaasalainen et al. 2003; Demura et al. 2006).

We obtained two visible spectra (0.4-0.92 $\mu$m) of 1993 HA,
in November and December 2015 (Fig.~3). 
The latter was cut at around 0.87 $\mu$m because affected by a poor sky subtraction longward.
The two spectra are in very good agreement
up to $\sim$0.7~$\mu$m, while some differences arise at longer wavelengths.
The two spectra refer to almost opposite rotational phases (cf. Table~1), 
hence small heterogeneities on the surface cannot be excluded.
The spectral slope in the 0.45-0.7 $\mu$m range is
1.00$\pm$0.03 $\mu$m$^{-1}$ and 0.90$\pm$0.03 $\mu$m$^{-1}$ for the spectra acquired in
November and December 2015, respectively.
These values are in line with those of D-type Trojan asteroids
measured by Fornasier et al. (2004, 2007).
For comparison, the reddest slope measured in the same wavelength range by Neely et al. (2014) for
metallic asteroids are 0.67 $\mu$m$^{-1}$
for (16) Psyche and 0.76 $\mu$m$^{-1}$ for (441) Bathilde.
This makes a classification in the X-type spectral group
(whose reddest components are the metallic asteroids)
unlikely,
and suggests a primitive nature for 1993 HA.
In Fig.~3 we also plot the visible components
of the D, T and X spectral types defined by DeMeo et al. (2009). 
We stress that such taxonomy is for objects with visible and near-infrared spectra,
but here we use it just for visual comparison, also relying on the fact
that the three plotted taxa are basically unchanged  
with respect to Bus' visible wavelength taxonomy (Bus 1999; Bus \& Binzel 2002).

We note that Mueller et al. (2011) have measured the size (0.337$^{+0.097}_{-0.078}$ m)
and albedo (0.140$^{+0.110}_{-0.077}$) of 1993 HA within the ExploreNEOs survey
(Trilling et al. 2010),
based on observations with NASA's Warm-Spitzer space telescope.
Such moderate albedo value
seems marginally compatible with a D-type classification (typical albedo $\la$ 0.10,
e.g., Ryan \& Woodward 2010; Mainzer et al. 2011),
while more in agreement with a T- or X-type.
However it is important to consider the large error bar,
as well as the fact that Harris et al. (2011) 
checked the accuracy of the ExploreNEOs results against
values published in the literature and found albedos to be typically consistent only within 50\%.
Hence new measurements of the albedo of 1993 HA would be extremely desirable.

The D-type classification seems to be reinforced when we take into account the
near-infrared spectrum we obtained ($\sim$0.92-1.58 $\mu$m), which has been
normalized to the visible data by assuming a linear behaviour in the 0.85-1.1~$\mu$m range
and scaling the NIR data accordingly. 
We used the M4AST online tool\footnote{http://m4ast.imcce.fr/} (Popescu et al. 2012),
and in particular the chi-square minimization method, 
to classify our complete visible and NIR spectrum
within the taxonomic scheme by DeMeo et al. (2009).
The featureless, red-sloped overall spectral shape of 1993 HA
is most indicative of a D-type asteroid ($\chi^2$=0.00731),
while T-type ($\chi^2$=0.00840) and X-type ($\chi^2$=0.01026) classes seem less compatible
with the observed spectrum (Fig.~4). 


   \begin{figure}
   \centering
   \includegraphics[width=9cm]{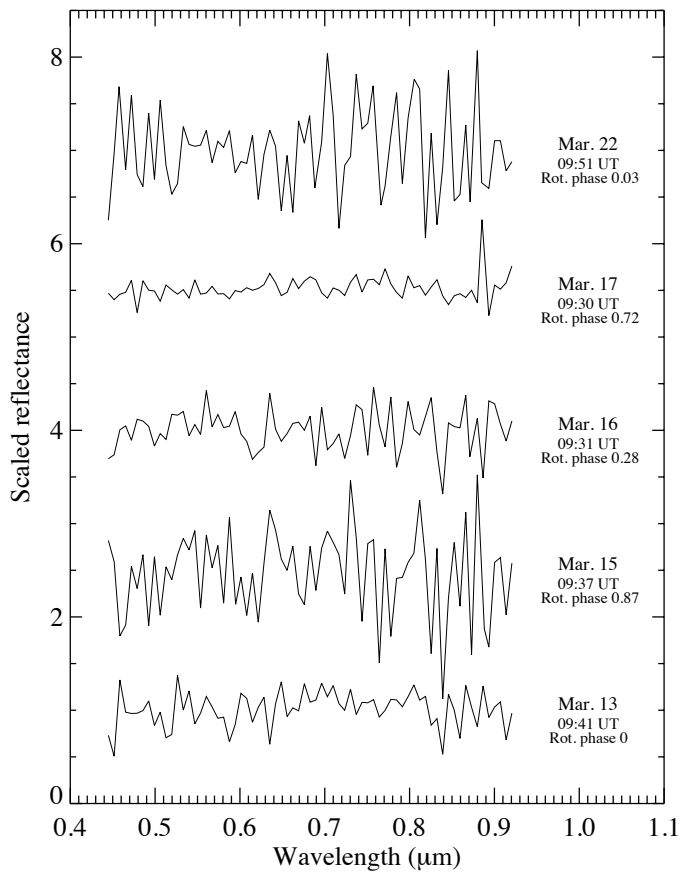}
      \caption{Visible spectra of 2008 EV5, acquired with the FORS2 instrument at the ESO-VLT.
      Spectra are normalized at 0.55 $\mu$m and shifted by 1.5 in reflectance for clarity.
      The mid-exposure time is reported for each spectrum,
      together with the corresponding rotational phase (P$_{rot}$=3.725$\pm$0.001 h).
      We remind the reader of the low S/N of these data (see Section 3.1):
      the spectra presented here are flat within the limits of noise.}
         \label{fig_EV5}
   \end{figure}

      \begin{figure}
   \centering
   \includegraphics[width=9cm]{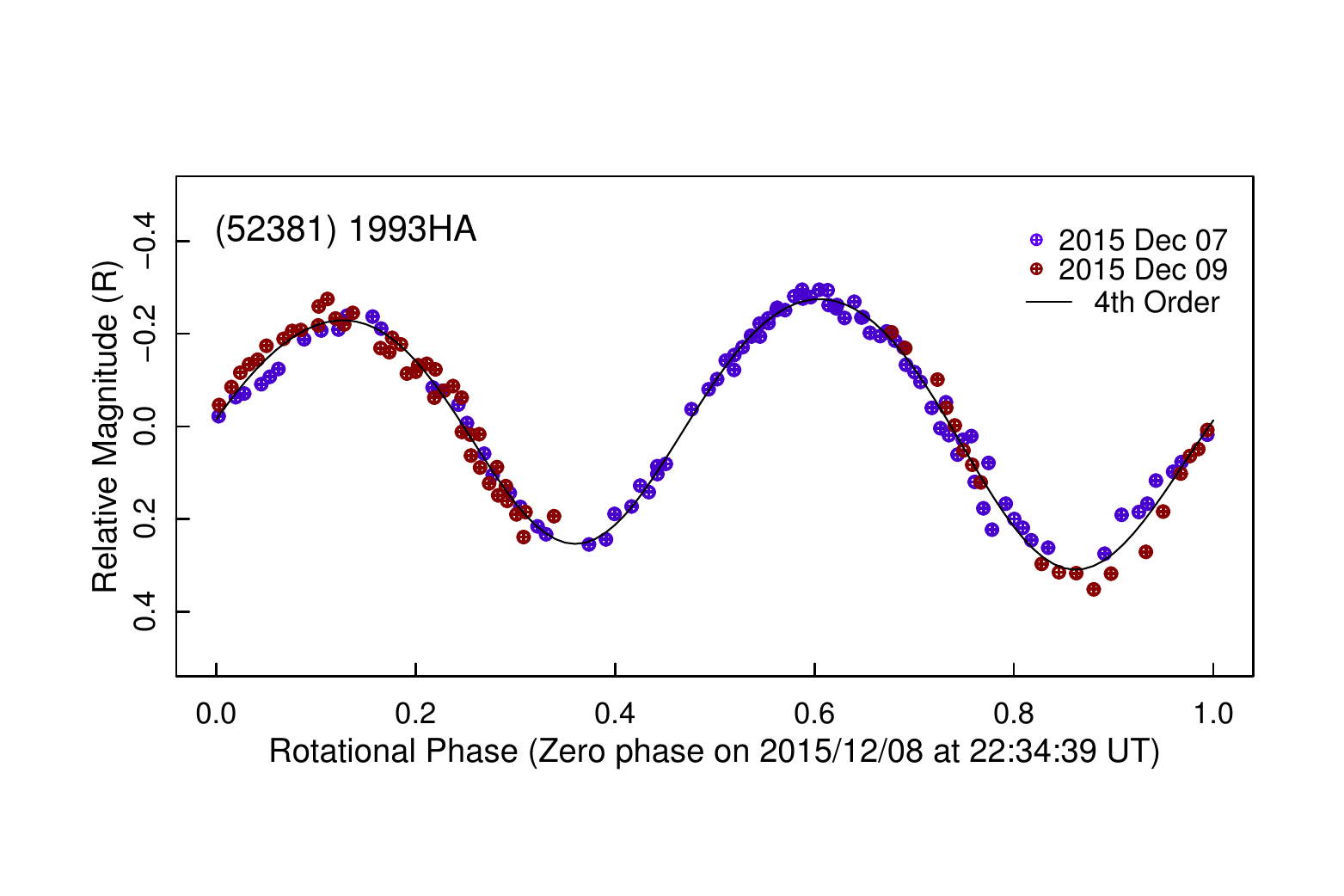}
      \caption{Composite lightcurve of 1993 HA, folded with a synodic period of 4.107 h.
      The average error on each photometric point is of $\sim$0.015 mag.}
         \label{fig_HA_LC}
   \end{figure}


   \begin{figure}
   \centering
  \includegraphics[width=9cm]{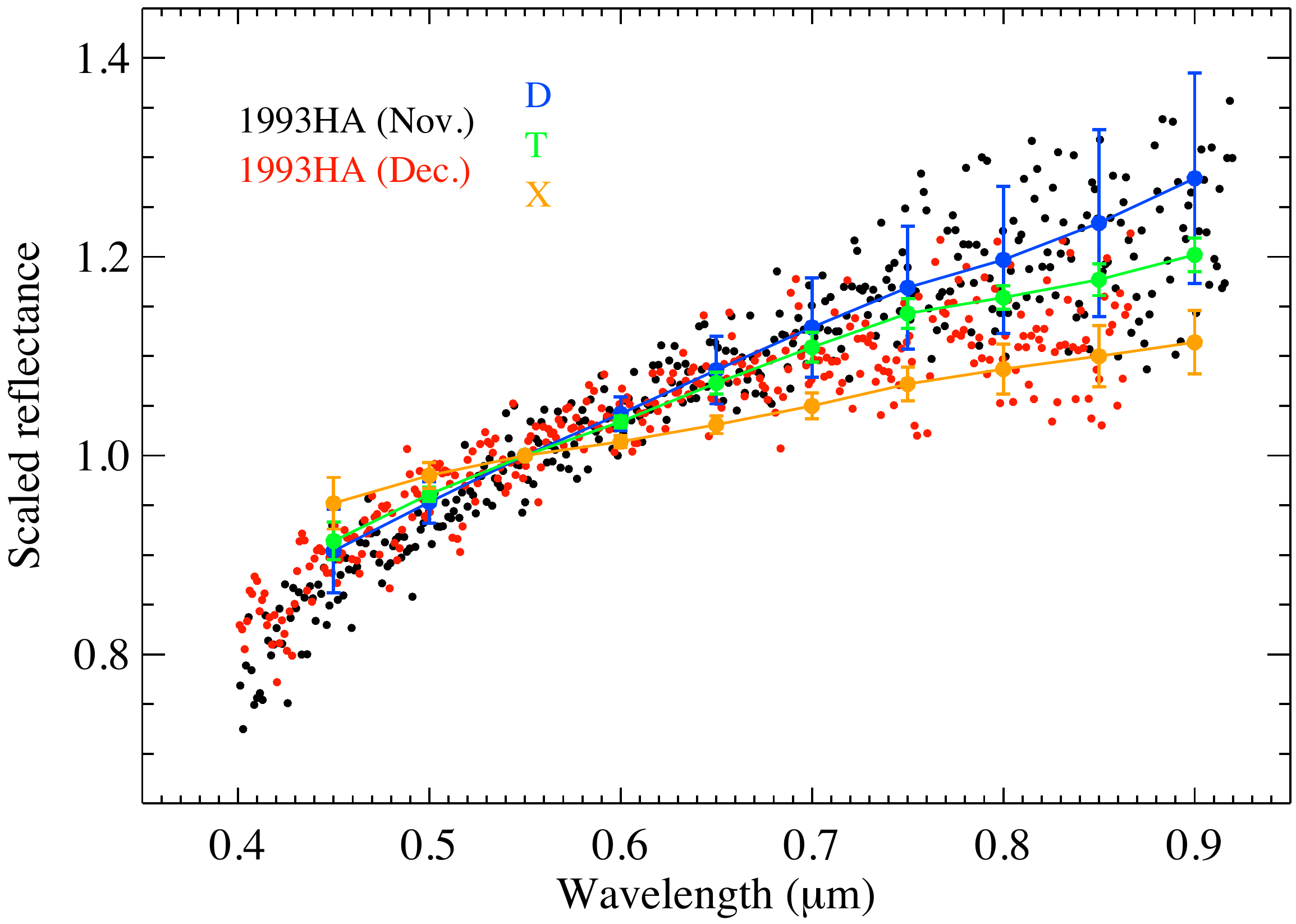}
      \caption{Visible spectra of 1993 HA (normalized at 0.55 $\mu$m),
      acquired with the EFOSC2 instrument at the ESO-NTT.
      The visible components of the
      average spectra of D-, T- and X-type asteroids from the DeMeo et al. (2009)
      taxonomic classification are also reported.}
         \label{fig_HA_vis}
   \end{figure}
   
      \begin{figure}
   \centering
   \includegraphics[width=9cm]{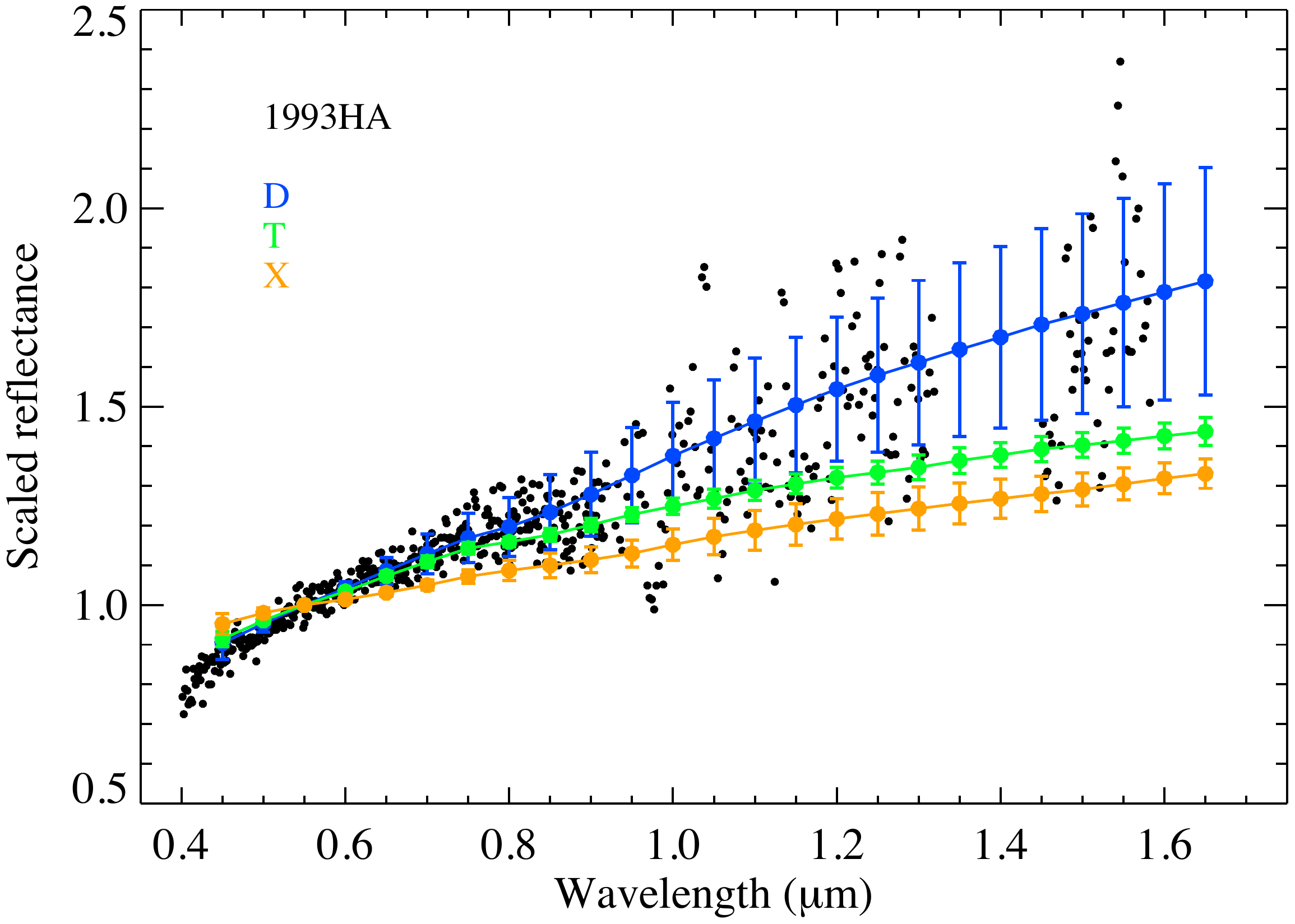}
      \caption{Combined visible and near-infrared spectrum of 1993 HA
      (EFOSC2 + SOfI at ESO-NTT).
      The average spectra of D-, T- and X-type asteroids from the DeMeo et al. (2009)
      taxonomic classification are also reported.}
         \label{fig_HA_visnir}
   \end{figure}


\section{Conclusions}
In this work we present new observations of the low-$\Delta$V
near-Earth asteroids (341843) 2008 EV5 and (52381) 1993 HA.

The asteroid 2008 EV5 is the baseline target of the NASA Asteroid Redirect Mission project.
It has been chosen
mainly based on its favourable orbital properties
($\Delta$V=5.6 km/s following the approach by Shoemaker \& Helin 1978)
and C-type spectral type.
However only one spectrum of this asteroid exists in the literature,
and it is important to confirm such spectral classification
before investing several hundred million euros on a spacecraft mission.
We note that the asteroid 2008 EV5 will not be observable again until late 2020, when
the final decision about the ARM mission target will have already been made.
Our new data agree with the C-type classification of 2008 EV5 and provide
evidence of the homogeneity of its surface through the
first rotational study of this asteroid.

The asteroid 1993 HA has never been spectrally characterized prior to our observations:
our visible and near-infrared data suggest that this object is a rare D-type near-Earth asteroid.
Although they are less probable, a T- or X-type classification cannot be completely ruled out,
especially considering the moderate albedo value, measured roughly by Warm-Spitzer.
For 1993 HA we also found a synodic rotation period of 4.107$\pm$0.002~h,
an optimal value for the execution of a sample return mission.
The lightcurve amplitude of 1993 HA also implies a quite elongated shape,
with a derived lower limit for the axis ratio
of the triaxial ellipsoid shape of a/b=1.71.
Considering its very low $\Delta$V (5.3 km/s)
and short mission duration (3.6 years), 1993 HA seems an ideal target for the
MarcoPolo-M5 sample return mission proposed in 2016 in the context of
the ESA M5 mission opportunity.
New observations of this object (it will be as bright as V=19.2 mag in April 2019)
will, however, be very welcome
to further investigate its physical properties.

\begin{acknowledgements}
We thank the anonymous referee for the constructive remarks on the manuscript.
D.P. and M.P. acknowledge financial
support from the NEOShield-2 project, funded by the European
Commission's Horizon 2020 program (contract No. PROTEC-2-2014-640351).
F.M. thanks CAPES for its fellowship and D.L. has
been supported by CNPq and FAPERJ (grants 305369/2009-1 and
E-26/201.213/2014).
\end{acknowledgements}

%
%


Abell, P. A., Mazanek, D. D., Reeves, D. M., et al. 2016, LPI, 47, 2217

~

Appenzeller, I., Fricke, K., Fürtig, W., et al. 1998, ESO Messenger, 94, 1

~

Barucci, M. A., et al. 2009, Marco Polo assessment study report, http://sci.esa.int/marco-polo/46019-marco-polo-assessment-study-report/

~

Barucci, M. A., Dotto, E., \& Levasseur-Regourd, A. C. 2011, A\&AR, 19, 48

~

Barucci, M. A., Cheng, A. F., Michel, P., et al. 2012, ExA, 33, 645

~

Binzel, R. P., Perozzi, E., Rivkin, A. S., et al. 2004a, M\&PS, 39, 351

~

Binzel, R. P., Rivkin, A. S., Stuart, J. S., et al. 2004b, Icarus, 170, 259

~

Bus, S. J. 1999, Compositional structure in the asteroid belt: Results of a spectroscopic survey,
PhD thesis, Massachusetts Institute of Technology

~

Bus, S. J., \& Binzel, R. P. 2002, Icarus, 158, 146

~

Busch, M. W., Ostro, S. J., Benner, L. A. M., et al. 2011, Icarus, 212, 649

~

Buzzoni, B., Delabre, B., Dekker, H., et al. 1984, ESO Messenger, 38, 9

~

DeMeo, F. E., Binzel, R. P., Slivan, S. M., \& Bus, S. J. 2009, Icarus, 202, 160

~

Demura, H., Kobayashi, S., Nemoto, E., et al. 2006, Science, 312, 1347

~

Fornasier, S., Dotto, E., Marzari, F., et al. 2004, Icarus, 172, 221

~

Fornasier, S., Dotto, E., Hainaut, O., et al. 2007, Icarus, 190, 622

~

Galad, A., Vilagi, J., Kornos, L., \& Gajdos, S. 2009, Minor Planet Bull., 36, 116

~

Harris, A. W., Young, J. W., Bowell, E., et al. 1989, Icarus, 77, 171

~

Harris, A. W., Mommert, M., Hora, J. L., et al. 2011, AJ, 141, 75

~

Ieva, S., Dotto, E., Perna, D., et al. 2014, A\&A, 569, A59

~

Kaasalainen, M., Kwiatkowski, T., Abe, M, et al. 2003, A\&A, 405, L29

~

Lazzaro, D. 2010, BAAA, 53, 315

~

Lu, E. T., \& Love,  S. G. 2005, Nature, 438, 177

~

Mainzer, A., Grav, T., Bauer, J., et al. 2011, ApJ, 743, 156

~

Moorwood, A., Cuby, J.-G., \& Lidman, C. 1998, ESO Messenger, 91, 9

~

Mueller, M., Delbo', M., Hora, J. L., et al. 2011, AJ, 141, 109

~

Neeley, J. R., Clark, B. E., Ockert-Bell, M. E., et al. 2014, Icarus, 238, 37

~

Perna, D., Ka\v{n}uchov\'{a}, Z., Ieva, S., et al. 2015, A\&A, 575, L1

~

Perna, D., Dotto, E., Ieva, S., et al. 2016, AJ, 151, 11

~

Perozzi, E., Rossi, A., \& Valsecchi, G. B. 2001, P\&SS, 49, 3

~

Popescu, M., Birlan, M., Binzel, R., et al. 2011, A\&A, 535, A15

~

Popescu, M., Birlan, M., \& Nedelcu, D. A. 2012, A\&A, 544, A130

~

Reddy, V., Le Corre, L., Hicks, M., et al. 2012, Icarus, 221, 678

~

Ryan, E. L., \& Woodward, C. E. 2010, AJ, 140, 933

~

Sanchez, J.-P. \& McInnes, C. R. 2013, in Asteroids. Prospective Energy and Material Resources, ed. V. Badescu (Springer, Berlin), 439

~

Shoemaker, E. M., \& Helin, E. F. 1978, NASACP, 2053, 245

~

Somers, J. M., Hicks, M., Lawrence, K., et al. 2010, BAAS, 42, 1055

~

Trilling, D. E., Mueller, M., Hora, J. L., et al. 2010, AJ, 140, 770

~

Ye, Q. 2011, AJ, 141, 32


\end{document}